\documentclass{nhm}
\usepackage{amsmath}
\usepackage{paralist}
\usepackage{graphics} 
\usepackage{epsfig} 

\usepackage{hyperref} 

\def\currentvolume{X}
 \def\currentissue{X}
  \def\currentyear{200X}
   \def\currentmonth{XX}
    \def\ppages{X--XX}

\theoremstyle{definition}

\title[K-core decomposition of Internet graphs]
{K-core decomposition of 
Internet graphs: hierarchies, self-similarity
and measurement biases.}

\author[J. I. Alvarez-Hamelin and A. Barrat and L. Dall'Asta and A. Vespignani]{}


 \email{alexv@indiana.edu}

\begin{document}
\maketitle

\centerline{\scshape Jose Ignacio Alvarez-Hamelin }
\medskip
{\footnotesize
\centerline{CONICET and Facultad de Ingenier\'{\i}a, Universidad de Buenos Aires,}
\centerline{Paseo Col\'on 850, C1063ACV Ciudad de Buenos Aires, Argentina}
} 

\medskip

\centerline{\scshape Luca Dall'Asta }
\medskip
{\footnotesize
 \centerline{Abdus Salam International Center for Theoretical Physics,}
  \centerline{Strada Costiera 11, 34014 Trieste, Italy}
} %

\medskip

\centerline{\scshape Alain Barrat}
\medskip
{\footnotesize
 \centerline{LPT (UMR du CNRS 8627), Universit\'e de Paris-Sud, France, and}
\centerline{Complex Networks Lagrange Laboratory, ISI, Torino 10133, Italy}
} %

\medskip

\centerline{\scshape Alessandro Vespignani}
\medskip
{\footnotesize
 \centerline{School of Informatics, Indiana University, Bloomington, 
IN 47408, USA, and}
\centerline{Complex Networks Lagrange Laboratory, ISI, Torino 10133, Italy}
}

\bigskip

 \centerline{(Communicated by the associate editor name)}

\begin{abstract} 
We consider the $k$-core decomposition of network models and 
Internet graphs at the autonomous system (AS) level. 
The $k$-core analysis allows to characterize networks
beyond the degree distribution and uncover structural properties and
hierarchies due to the specific architecture of the system. 
We compare the $k$-core structure obtained for AS graphs 
with those of several network models and
discuss the differences and similarities with the real 
Internet architecture. The presence of biases and the incompleteness of
the real maps are discussed and their effect on the $k$-core analysis is
assessed with numerical experiments simulating biased exploration on a
wide range of network models. We find that the $k$-core analysis
provides an interesting characterization of the fluctuations and
incompleteness of maps as well as information helping to discriminate
the original underlying structure. 
\end{abstract} 

\maketitle 
 
\section{Introduction}\label{intro} 

In recent times, mapping projects of the World Wide Web (WWW) and the
physical Internet have offered the first chance to study topology and
traffic of large-scale networks. The study of large
scale networks, however, faces us with an array of new challenges. The
definitions of centrality, hierarchies and structural organizations
are in particular 
hindered by the large size of the systems and the complex
interplay of engineering, traffic, geographical, and
economical attributes characterizing their construction. 

In this paper we propose the $k$-core decomposition as a graph analysis
tool able to highlight interesting structural properties that are not
captured by the degree distribution or other simple topological
measures.The $k$-core
decomposition~\cite{Seidman,BollobaThomason83,Batagelj02} consists in
identifying particular subsets of the network, called $k$-cores, each
one obtained by a recursive pruning strategy. 
The $k$-core decomposition therefore provides a
probe to study the hierarchical properties of large scale networks,
focusing on the network's regions of increasing centrality and
connectedness properties. More central cores are indeed more strongly
connected, with larger number of possible distinct paths between
vertices: this allows to obtain not only more robust routing
properties but also a better opportunity to find a path with specific Quality of Service (QoS).

Here we study a set of basic network models and the AS level Internet 
maps obtained in two large scale measurement projects using very 
different techniques. We first characterize the $k$-core structure of
real Internet maps and compare with the structure obtained in
the various models. We find that the $k$-core structure is extremely
different in light tailed and heavy-tailed networks and is able to
clearly discriminate among various models presented in the literature. In
this perspective the $k$-core analysis represents a useful tool in the
model validation process. Moreover, any result concerning Internet maps has to 
consider their incompleteness and
the presence of measurements biases. For this reason we also present a
study of the stability of the $k$-core analysis in the presence of
biases and incomplete sampling in all the network models
considered. Our findings indicate that the $k$-core
decomposition's fingerprints allow the discrimination between
heterogeneous and homogeneous topologies even after an incomplete
sampling: this shows that the signatures observed
in the AS Internet maps are qualitatively reliable, even if some
biases are unavoidable at a detailed quantitative level.

\section{Related work}\label{related} 

In the last years, a wealth of studies have focused on the large scale
structure and heterogeneities of networked structure 
of practical interest in social science, critical infrastructures and
epidemiology~\cite{Barabasi:2000,mdbook,psvbook}.  
The Internet has been readily considered as a prototypical example of
complex network by the scientific community and starting with the
seminal paper by Faloutsos, Faloutsos and Faloutsos~\cite{falou99} an
impressive number of papers has dealt with the characterization of its
large scale properties and
hierarchies~\cite{crit_int2,gov02,psvbook,Subramanian,caida05,LAWD04}.
While the initial interest has been focused on the general principles
leading to the basic organization features of complex networks, the
research activity is now diving into system specific features that
distinguish and highlight the various forces and/or engineering at
work in each class of networks. This is a particular pressing need in
the Internet where even at the Autonomous System (AS) level 
the large scale self-organization 
principles are working along with economical and technical constraints, 
optimization principles and so on~\cite{LAWD04,doylewillpnas}.
In addition actual Internet maps are not free from errors and
measurement biases. For this reason, 
recent works have been devoted to a better understanding of 
the possible sources of errors and biases presented by the
experimental
data~\cite{crovella,delos04,traceorsay,traceorsay2,latapy,claus05,viger}.
Since Internet maps are typically based on a sampling of routes between
sources and destinations (obtained by tools such as \texttt{traceroute}), 
these studies have dealt with simplified models of 
\texttt{traceroute}-like sampling, applied to 
graphs with various topological properties. They have shown that,
except in some peculiar cases~\cite{claus05}, the sampling process allows to
distinguish qualitatively between networks with strongly different
properties (homogeneous vs. heterogeneous), while a quantitative and
detailed view of the network may suffer important biases. 

Here, we consider the use of the $k$-core decomposition as a probe for
the structure of Internet maps. The $k$-core decomposition has 
mostly been used in biologically related contexts, where it was 
applied to the analysis of protein
interaction networks or in the prediction of protein functions
\cite{bader03,Wuchty05}.  An interesting application in the area of
networking has been provided by Gkantsidis 
{\em et al.}~\cite{Gkantsidis} and Gaertler {\em et
al.}~\cite{Gaertler04}, where the $k$-core decomposition is used for
filtering out peripheral Autonomous Systems (ASes) in the case of
Internet maps. The $k$-core decomposition has also recently been used
as a basis for the visualization of large networks, in particular for
AS maps~\cite{baur04,ignacio05,LANET-VI}.  Finally, recent works using
the $k$-core analysis have focused on the analysis of the
Internet maps obtained by the DIMES project~\cite{DIMES}. In
Ref.s~\cite{medusa,medusa2}, an approach based on the $k$-core
decomposition has been used to provide a conceptual and structural
model of the Internet, the so-called Medusa model for the Internet.
Up to now, no study has however considered the $k$-core decomposition
of the various commonly used models for complex networks, nor compared
it to the one of real-world networks. 
Subramanian {\em et al.}~\cite{Subramanian} have proposed to classify
ASes in five different levels or "tiers", and given a method to
extract this classification from the AS directed graph. This method
can however lead to some biases when the knowledge of the all
peer-to-peer relationships is not complete. The $k$-core
decomposition studied in this paper considers on the other hand
undirected networks, and yields a finer hierarchy, not based on the
commercial relations between vertices, and in which the number of
levels is not fixed a priori but depends on the characteristics of the
network. It is moreover not restricted to AS maps but can be applied
as well for example to Internet router maps or more generally to any
real or computer generated graph.

\section{{\large $k$}-core decomposition}\label{sec:k-core} 
 
Let us consider a graph $G=(V,E)$ of $|V|=n$ vertices and $|E|=e$ edges, the
definition from~\cite{Batagelj02} of $k$-cores is the following

{\bf Definition 1}: A subgraph $H = (C,E|C)$ induced by the set
$C\subseteq V$ is a {\em $k$-core} or a core of order $k$ if and only
if the degree of every node $v \in C$ induced in $H$ is greater or equal
than $k$ (in symbolic form, this reads 
$\forall v \in C: {\tt degree}_H(v)\geq k$), and $H$ is the maximum
subgraph with this property.

A $k$-core of $G$ can therefore be obtained by recursively removing all the
vertices of degree less than $k$, until all vertices in the remaining graph
have degree at least $k$. It is worth remarking that this process is
not equivalent to prune vertices of a certain degree. Indeed, a
star-like subgraph formed by a vertex with a high degree that connects many
vertices with degree one, and connected only with a single edge to the
rest of the graph, is going to belong to the first shell no matter how
high is the degree of the vertex. We will also use the following definitions

{\bf Definition 2}: 
A vertex $i$ has {\em shell index} $k$ if it belongs to the 
$k$-core but not to $(k+1)$-core.

{\bf Definition 3}: 
A {\em $k$-shell} $S_k$ is composed by all the vertices whose shell
index is $k$.  The maximum value $k$ such that $S_k$ is not empty is
denoted $k_{\max}$. The $k$-core is thus the union of all shells $S_c$
with $c \ge k$.

{\bf Definition 4}:
Each connected set of vertices having the same shell index $c$ 
is a {\em cluster} $Q^c$, where the corresponding set of edges are those
connecting vertices of the cluster.
Each shell $S_c$ is thus composed by clusters $Q_m^c$, such that 
$S_c=\cup_{1\leq m \leq q_{\max}^c} Q^c_m$, where $q_{\max}^c$ is 
the number of clusters in $S_c$.

\begin{figure}[t] 
\begin{center} 
\includegraphics[width=5.1cm,angle=-90]{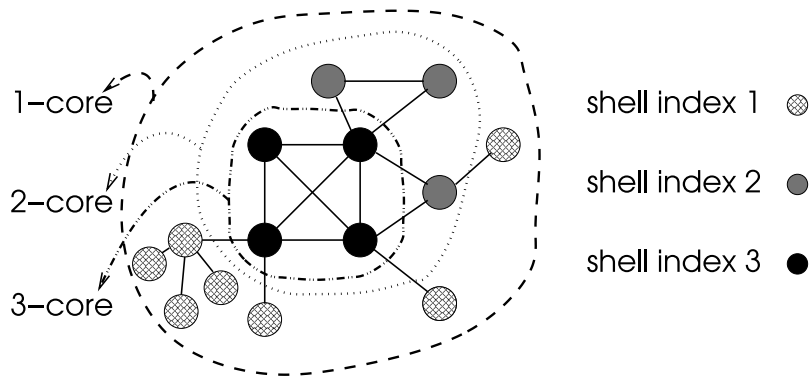} 
\end{center} 
\caption{Sketch of the $k$-core decomposition for a small
graph. Each closed line contains the set of vertices belonging to a
given $k$-core, while different types of vertices correspond to different
$k$-shells.}
\label{k-core1}
\end{figure}

The $k$-core decomposition therefore identifies progressively internal
cores and decomposes the networks layer by layer, revealing the
structure of the different $k$-shells from the outmost one to the most
internal one, as sketched in Fig.~\ref{k-core1}.

It is worth to note that the $k$-core decomposition can be
easily implemented: the algorithm by Batagelj and
Zversnik~\cite{Batagelj03} presents a time complexity of order 
$O(n+e)$ for a general graph. This makes the algorithm
very efficient for sparse graphs, where $e$ is of order $n$.

A very interesting feature of the $k$-cores concerns their
connectivity properties. It has been for example shown experimentally
in~\cite{medusa2} that the $k$-cores of the AS map obtained by the
DIMES project~\cite{DIMES} are $k$-connected, which means that $k$
disjoint paths are available between any two vertices belonging to the
$k$-core. In fact, for any two vertices $u$ and $v$ of the network,
with shell indices respectively $c_u$ and $c_v$, there are (with some
exceptions for small values of $c_u$ and $c_v$) at least
$\min(c_u,c_v)$ disjoint paths between $u$ and $v$ \cite{medusa2}. 
Such property has important practical consequences since it implies
larger and larger robustness and routing capacities for more and more
central cores. The knowledge of such capacities identifies a very
important hierarchy of ASes that could be taken advantage of by newly
created ASes in order to choose to which other ASes to establish
connections. We will come back to this point in section \ref{sec:appl}.

\section{{\large $k$}-core structure of Internet maps and models} 
\label{casestudy} 

\subsection{Internet AS maps}

In this section, we inspect Internet maps at the AS level and compare
their $k$-core structure with the insights obtained from models.  In
order to obtain Internet connectivity information at the AS level it
is possible to inspect routing tables and paths stored in each router
(passive measurements) or directly ask the network with a software
probe (active measurements). In the following we consider data from
two recent large scale Internet mapping projects using an active
measurement approach.  The skitter project at CAIDA~\cite{IR_CAIDA}
has deployed several strategically placed probing monitors using a
path probing software.  All the data are then centrally collected and
merged in order to obtain Internet maps that maximizes the estimate of
cross-connectivity. The second set we consider is provided by the
Distributed Internet Measurements and Simulations (DIMES)
project~\cite{DIMES,shavitt}. At the time where the map was obtained,
the project consisted of more than 5,000 measuring agents performing
Internet measurements such as \texttt{traceroute} and \texttt{ping}.
Table~\ref{table2} displays a summary of the basic properties of the
considered Internet maps. 
We have also investigated Internet maps obtained from the Oregon Routeviews
project \cite{oregon} (not shown), with very similar results.
In the following we show how the application
of the $k$-core decomposition can shed light on important hierarchical
properties of Internet graphs, focusing on the AS maps obtained by
each project in 2005.

\begin{figure}[hbt]
\begin{center}
\includegraphics[width=8.1cm,angle=0]{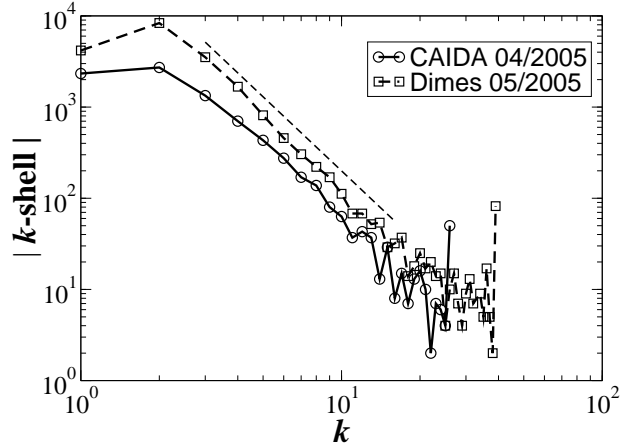}
\end{center}
\caption{Shell size as a function of their index for the AS maps. The
dashed line is a power-law $\propto k^{-2.7}$.}
\label{shell_AS}
\end{figure}

The first observation about the structure of the $k$-cores is that
they remain connected. This is not a
priori an obvious fact since one can easily imagine networks whose
$k$-core decomposition yields several connected components
corresponding, {\em e.g.} to various communities. Instead, each
decomposition step is just {\em peeling} the network leaving connected
the inner part of the network, showing a high hierarchical structure,
{\em i.e.} the most connected part of the network is also the most
central. Figure~\ref{shell_AS} displays the size in terms of vertices
of each $k$-shell as a function of its index. As for RSF or BRITE
networks (see section~\ref{sec:models}), 
power-law like shapes are obtained. Important
fluctuations appear at large $k$, which is not very surprising since
such shells of large index are relatively small, except for the most
central core which contains $50$ vertices at $k_{max}=26$ and
$82$ vertices at $k_{max}=39$ for CAIDA and DIMES, respectively. 
Such a structure has
also been observed in the independent study of~\cite{medusa2}.

\begin{table}[thb]
\small
\begin{tabular}{|c|r|r|c|c|c|}
\hline
source  & n & e & $\langle d\rangle$& $d_{max}$& $k_{\max}$ \\
\hline  
CAIDA, 2005/04                    & 8542  & 25492 & 5.97 & 1171 & 26 \\
\hline 
DIMES, 2005/05                    & 20455 & 61760 & 6.04 & 2800 & 39 \\
\hline
\end{tabular}
\caption{Main properties of the Internet maps considered in the
present study: number of vertices $n$ and of edges $e$, average degree
$\langle d \rangle$, maximum degree $d_{max}$ and maximum shell
index $k_{\max}$.}
\label{table2}
\end{table}

Interestingly, a much larger $k_{max}$ is obtained for the DIMES AS
map than for the CAIDA one.  It is likely that such discrepancy is
linked to the diversity of the exploration methods. The maximum core
depends indeed largely on the amount of discovered edges and lateral
connectivity. The set of ``observers'' is $22$ for CAIDA but more than
$5,000$ for DIMES. It is therefore reasonable that the latter has more
probability to discover edges, and therefore a larger value of
$k_{\max}$.

\subsubsection{Self-similarity}

The properties of the successive $k$-cores of Internet maps can be studied by
considering their degree distributions and correlation properties.

\begin{figure}[thb]
\begin{center}
\includegraphics[width=9.1cm,angle=0]{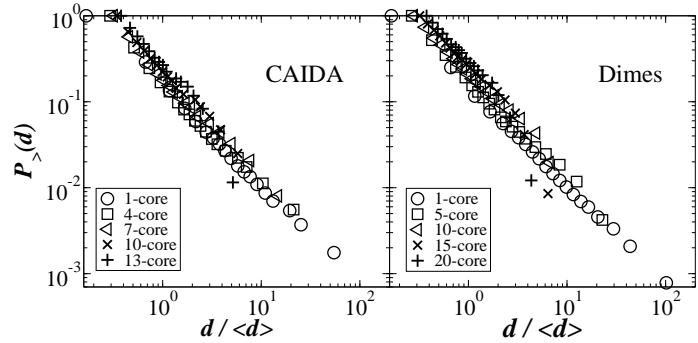}
\end{center}
\caption{Rescaled cumulative degree distributions of some $k$-cores
of the AS Internet maps. The degree is normalized by the corresponding
average degree $\langle d \rangle$ 
in each $k$-core. The shapes of the distributions
are preserved by the successive pruning, pointing to a self-similar 
behavior of the successive $k$-cores.}
\label{Pk_AS}
\end{figure}

Figure~\ref{Pk_AS} shows the cumulative degree distribution for the first
$k$-cores, for the various AS maps. Strikingly, the shape of the distribution,
{\em i.e.} an approximate power-law, is not affected by the decomposition.
This is illustrated by the fact that the data for the various distributions
collapse on top of each other, once the degree is rescaled by the average
degree of the $k$-core. Note that in Fig.~\ref{Pk_AS}, as in the following
figures, we do not show data for {\em all} the cores, but only for a
representative set of $k$ values.
Figure~\ref{Pk_AS} clearly shows how the exponent of the power-law is robust
across the various $k$-cores,
although the range of variation of the degree decreases. In other words, each
core conserves a broad degree distribution: AS with significantly different
number of neighbors are present in each core or hierarchy level.

In order to better characterize and check this self-similarity, we
have computed also the two and three points correlations functions of
the various $k$-cores. A useful measure to quantify correlations
between the degrees of neighboring vertices is the {\em average degree
of nearest neighbors} $d_{nn}(d)$ of vertices of degree $d$
\cite{alexei}:
\begin{equation}
d_{nn}(d)=\frac{1}{n_d}\sum_{j / d_j=d}
\frac{1}{d_j} \sum_{i \in V(j)} d_i \quad ,
\end{equation}
where $V(j)$ is the set of the
$d_j$ neighbors of vertex $j$ and $n_d$ the number
of vertices of degree $d$.
This last quantity is related to the correlations between the degree
of connected vertices since on the average it can  be  expressed as
\begin{equation}
d_{nn}(d) = \sum_{d'} d' P(d'|d),
\end{equation}
where $P(d'|d)$ is the conditional probability that a vertex with
degree $d$ is connected to a vertex with degree $d'$.  If degrees of
neighboring vertices are uncorrelated, $P(d'|d)$ depends only on
$d'$ and thus $d_{nn}(d)$ is a constant. When correlations are
present, two main classes of possible correlations have been
identified: {\em assortative} behavior if $d_{nn}(d)$ increases with
$d$, which indicates that large degree vertices are preferentially
connected with other large degree vertices, and {\em disassortative}
if $d_{nn}(d)$ decreases with $d$~\cite{Newman:2002}.  From a routing
point of view, a disassortative behavior corresponds to a network
structure where vertices with small degree are
preferentially connected to the hubs (i.e.,
large degree vertices). A second, and
often studied, relevant quantity is the {\em clustering coefficient}
\cite{watts98}
that measures the local group cohesiveness and is defined for any
vertex $j$ as the fraction of connected neighbors of $j$
\begin{equation}
cc_j=2\cdot n_{\tt link} / (d_j(d_j-1))\enspace,
\end{equation}
where $n_{\tt link}$ is the number of links between the $d_j$
neighbors of $j$. The study of the clustering spectrum $cc(d)$ of
vertices of degree $d$, defined as
\begin{equation}
cc(d)= \frac{1}{n_d}\sum_{j / d_j=d} cc_j \quad ,
\end{equation}
allows, {\em e.g.} to uncover hierarchies in
which low degree vertices belong generally to well interconnected
communities (high clustering coefficient), while hubs connect many
vertices that are not directly connected (small clustering
coefficient). Large clustering has a clear relevance for routing
purposes since it indicates the presence of alternative paths thanks
to the presence of many triangles: if a link from a vertex $u$ to a
neighbor $v$ goes down, the message can be sent from $u$ to $v$
through a common neighbor.

\begin{figure}[thb]
\begin{center}
\includegraphics[width=9.1cm,angle=0]{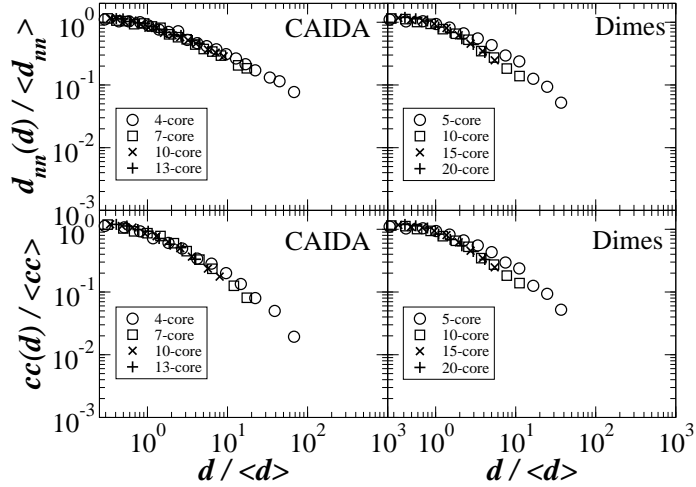}
\end{center}
\caption{Average nearest neighbor (top) and rescaled clustering
  spectrum (bottom) as a function of the degree for some $k$-cores of
  the AS Internet maps. All the quantities are rescaled by the corresponding
  averages in each $k$-core. The collapse of the various curves confirm
 the self-similar structure of the $k$-cores.}
\label{KnnCnn_AS}
\end{figure}

Figure \ref{KnnCnn_AS} shows that not only the degree distribution
but also the clustering and correlations structures
of the Internet maps are essentially preserved as the more and more
external parts of the network are pruned. We note however that, as also shown
in \cite{medusa2}, the largest $k$-cores are no more scale-free: since they
are very densely connected, their degree distribution is rather peaked around
an average value and their topology is closer to that of a random graph with
large average degree.

In summary, the AS networks exhibit a statistical scale invariance
with respect to the pruning obtained with the $k$-cores decomposition
for a wide range of $k$. Indeed, while this decomposition identifies
subgraphs that progressively correspond to the most central regions of
the network, the statistical properties of these subgraphs are
preserved at many levels of pruning. This hints to a sort of global
self-similarity for regions of increasing centrality of the network,
and to a structure in which each region of the Internet as defined in
terms of network centrality has the same properties than the whole
network. This is particularly interesting since the properties of
Internet (heterogeneous degree distributions, correlations,
clustering...) have been up to now studied at the level of the whole
map, while one can be interested to restrict the analysis to some
particular regions of the map, focusing for example on parts of the
network with certain routing capabilities (QoS, failure support).
At a general level, the $k$-core decomposition appears
therefore as a suitable way to define a pruning procedure equivalent
to a scale-change preserving the statistical properties of graphs
while focusing on their more and more connected parts.

\subsubsection{Shell index and centrality}

The identification of the most central vertices is a major issue in
networks characterization~\cite{freeman77}.  While a first intuitive
and immediate measure of the centrality of vertices is given by their
degree, more refined investigations are needed in order to
characterize the real importance of various vertices: for example,
some low-degree vertices may be essential because they provide
connections between otherwise separated parts of the network. In order
to uncover such important vertices, the concept of betweenness
centrality (BC) is now commonly used
\cite{freeman77,newman01b}. The betweenness centrality of a
vertex $v$ is defined as
\begin{equation}
 g(v)=\sum_{s\neq t}\frac{\sigma_{st}(v)}{\sigma_{st}} \quad, 
\end{equation}
where $\sigma_{st}$ is the
number of shortest paths going from $s$ to $t$ and $\sigma_{st}(v)$ is
the number of shortest paths from $s$ to $t$ going through $v$. This
definition means that central vertices are part of more shortest paths
within the network than peripheral vertices.  Moreover, the
betweenness centrality gives in transport networks an estimate of the
traffic handled by the vertices, assuming that the number of shortest
paths is a zero-th order approximation to the frequency of use of a
given vertex ({\em e.g.} the load of an AS), in the case of an {\em
all-to-all} communication.

The $k$-core decomposition 
intuitively provides a hierarchy of the vertices based on their shell
index that is a combination of local and global properties. ({\em
e.g.},~\cite{Wuchty05} shows that the shell index is a better
criterium for centrality than the degree in protein interaction
networks).
\begin{figure}[t] 
\begin{center} 
\includegraphics[width=90mm,angle=0]{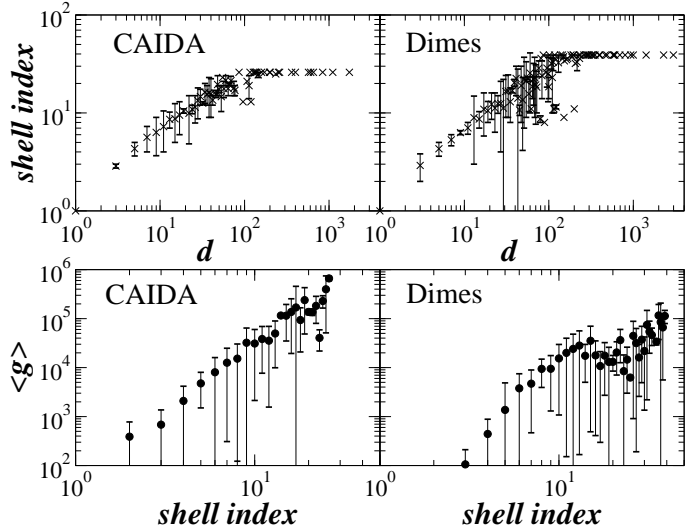} 
\end{center} 
\caption{Average betweenness centrality as a function of
shell index (bottom), and average shell index as a function of the degree
(top),
for the AS Internet maps. A clear correlation between these quantities
is observed, although strong fluctuations are present.}
\label{centrality} 
\end{figure} 
In this perspective, it becomes very interesting to study the correlation
between the degree, the betweenness centrality and the shell index of a vertex
in order to quantify the statistical level of consistency of the various
measures. We show in Fig.~\ref{centrality} the average betweenness centrality
(computed on the original graph) of vertices as a function of their shell
index, and the shell index as a function of the degree $d$. A strong
correlation is expected, but the fluctuations observed (given by the
errorbars) should not be a surprise: while a low-degree vertex has clearly low
shell index, large or medium degree vertices do not have necessarily a large
shell index. In the AS maps, we observe in fact that all large degree vertices
belong to the most central core, while large fluctuations are observed for
intermediate degree values.  Moreover, the betweenness centrality is a highly
non-local quantity which can be large even for small-degree vertices. These
quantities are thus pinpointing different kinds of centrality. The shell index
appears therefore as a very interesting quantity to uncover central vertices
and it has the advantage of a much faster computation time than those required
for the betweenness centrality (of order $n^2 \log n$~\cite{Brandes:2001}).

\subsubsection{Potential practical implications}
\label{sec:appl}

The $k$-core decomposition has interesting immediate applications.
First of all, as already mentioned in section~\ref{sec:k-core}, it has
been shown in ref~\cite{medusa2} that each $k$-core of the DIMES AS map
is $k$-connected, and that the number of disjoint paths between
two vertices $u$ and $v$ of this map is bounded from below by the minimum
of the shell indices of $u$ and $v$.

Moreover, it is quite easy to show and understand that similar
properties hold for a network under certain assumptions. In
particular, if the central core (of shell index $k_{max}$) of a given
network is $k_{max}$-edge-connected, and if there exists enough edges
between the various shells (in particular if any cluster -see Def. 4- of each
$k$-shell is connected to the $k+1$-shell by at least $k$ edges), then
each $k$-core of the network turns out to be $k$-edge connected.  We
have in fact checked that these conditions are verified for the CAIDA
and DIMES maps as well as for the network models under study.
Note that $k$-edge connectivity ({\em i.e.} the existence of $k$ distinct
paths which do not share any common edge) is less restrictive than
$k$-connectivity. In the context of Autonomous Systems and evaluation
of routing capacities or of failure possibilities however, it is
particularly relevant since a vertex of the AS map represents in fact
many different routers, so that different paths
may cross at a given AS while being effectively physically disjoints.

Such connectivity properties highlight the fact that the $k$-core
decomposition provides a natural definition for a hierarchy in the network,
in which the more central vertices (with larger shell index) have
better routing capabilities ({\em i.e.} they can choose several paths to
achive a certain connection), and each $k$-core constitutes an ensemble
of ASes able to provide a certain QoS, with global
larger robustness for larger $k$.

It is therefore interesting to compare the $k$-core decomposition with
the tiers hierarchy proposed by Subramanian et al.~\cite{Subramanian}.
These two hierarchies have different origins and motivations: on the
one hand, the tiers classification is based on the inference of AS
commercial relationships; on the other hand, and in a somehow opposite
point of view, the $k$-core decomposition gives a classification of
the network's vertices which does not have an a priori fixed number of
classes or levels, but which adapts itself to the situation of the
network. Moreover, the shell index of a vertex is not fixed once and
for all but may fluctuate in time due to possible connectivity changes
(as investigated in the next section). In this aspect, such a
hierarchy provides very relevant information about the state of the
network at a given time. While the actual routing protocols do not
take advantage of such information, one could imagine that future
routing protocols may be able to exploit it.

We finally note that the use of the $k$-core decomposition in order
to find a certain hierarchy of connectedness properties is not
limited to the analysis of AS maps: it can as well be applied
to other kinds of Internet maps, for example at the router level, 
or to any communication or transportation network.

\subsection{$k$-core structure of network  models}
\label{sec:models}
In order to better understand the properties of the $k$-core
decomposition of networks and use it as a model validation tool, we
also apply this technique to a set of well known and commonly used
models of networks, whose main characteristics are summarized in
Table~\ref{table1}. Various topological properties can lead to various
decompositions so we consider both homogeneous and heterogeneous
networks. For each model, we will present results corresponding to one random
instance of the model, and have checked that the highlighted properties
do not depend on the particular instance considered.

\begin{table}[thb]
\small
\begin{tabular}{|c|r|r|c|c|c|}
\hline
source  & n & e & $\langle d\rangle$& $d_{max}$& $k_{\max}$ \\
\hline
ER  & $10^5$ & $10^6$ & $20$ & $41$ & $14$ \\
\hline
BA& $5. 10^4$ & $99998$ & $4$ & $642$ & $3$ \\
\hline
Weibull & $10^5$ & $307500$  & $6.15$  &$377$ & $9$ \\
\hline
RSF $\gamma=2.3$ & $97315$ & $293891$ & $6.04$  & $938$  & $22$ \\
\hline
BRITE  & $10^5$ &  $156145$ & $3.63$ & $433$ & $54$ \\
\hline
INET 3.0 & $10000$ & $19676$ & $3.936$ &  $984$ & $8$ \\
\hline
\end{tabular} 
\caption{Main properties of the models considered in the
present study: number of vertices $n$ and of edges $e$, average degree
$\langle d \rangle$, maximum degree $d_{max}$ and maximum shell
index $k_{\max}$.}
\label{table1}
\end{table}

\subsubsection{Size of shells}

\begin{figure}[t] 
\begin{center} 
\includegraphics[width=9.cm,angle=0]{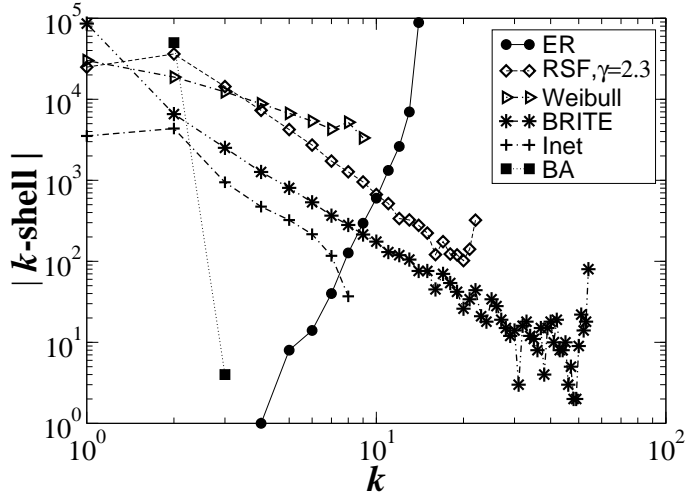} 
\end{center} 
\caption{Shell size as a function of their index for the various models
considered. The various models yield very different shapes, indicating
the $k$-core decomposition as an interesting additional tool for
network characterization.}
\label{fig1} 
\end{figure}

We first consider for reference the random graph model of Erd\"os and
R\'enyi (ER)~\cite{ErdosRenyi59}, which is the most standard example
of graphs with a characteristic value for the degree (the average
value $\langle d \rangle$). In this case, the maximum index is clearly
related to the average degree $\langle d\rangle$. The vertex degrees
have only small fluctuations, thus most vertices belong to the same
$k$-core that is also the highest. Noticeably, the size of the shells
is increasing with the index, showing that only few vertices can be
considered as peripherical (see Fig.~\ref{fig1}), and that the network
contains no clear hierarchy between nodes.

A second model we considered is the Barab\'asi-Albert (BA)
model~\cite{sf99} that has been put forward to exemplify the concept of
preferential attachment and as a paradigm of dynamically evolving networks. 
In this model, a growing network is constructed
according to the preferential attachment mechanism: each new vertex is
connected to $m$ already existing vertices chosen with a probability
proportional to their starting degree. This model produces graphs with
power-law degree distributions, thus characterized by a very large
variety of degree values. On the other hand, this is a toy model that
should not be considered as a realistic model in the Internet and
indeed the corresponding $k$-core
decomposition is somehow trivial, with only few shells at very small index.
The construction mechanism provides a simple explanation. Each new
vertex enters the system with degree $m$, but at the following time
steps new vertices may connect to it, increasing its degree. Inverting
the procedure, we obtain exactly the $k$-core decomposition. The
minimum degree is $m$, therefore all shells $C_c$ with $c < m$ are
empty. Recursively pruning all vertices of degree $m$, one first
removes the last vertex, then the one added at the preceding step,
whose degree is now reduced to its initial value $m$, and so on, up to
the initial vertices which may have larger degree. Hence, all vertices
except the initial ones belong to the shell of index $m$. 

Other algorithms are widely used to obtain
random graphs with prescribed broad degree distributions. In the
literature, different definitions of heavy-tailed like distributions
exist. While we do not want to enter the detailed definition, we have
considered two classes of such distributions: (i) {\em scale-free} or
Pareto distributions of the form $P(k) \sim k^{-\gamma}$ (RSF), and
(ii) Weibull distributions (WEI) $P(k)=(a/c) (k/c)^{a-1}
\exp(-(k/c)^a)$. The scale-free distribution has a diverging second
moment and therefore virtually unbounded fluctuations, limited only by
eventual size-cut-off. The Weibull distribution is akin to power-law
distributions truncated by an exponential cut-off which are often
encountered in the analysis of scale-free systems in the real
world. Indeed, a truncation of the power-law behavior is generally due
to finite-size effects and other physical constraints.  Both forms
have been proposed as representing the topological properties of the
Internet \cite{broido}. We have generated the corresponding random
graphs by using the algorithm proposed by Molloy and Reed
\cite{Molloy:1995,Molloy:1998}: the vertices of the graph are assigned
a fixed sequence of degrees $\{{k}_{i}\}$, \ $i = 1, \dots, N$, chosen
at random from the desired degree distribution $P(k)$, and with the
additional constraint that the sum $\sum_{i} {k}_{i}$ must be even;
then, the vertices are connected by $\sum_{i} {k}_{i} / 2$ edges,
respecting the assigned degrees and avoiding self- and
multiple-connections. The parameters used are $a=0.4$ and $c=0.6$ for
the Weibull distribution, and $\gamma=2.3$  for the RSF case.

The previous construction can be considered as static as
it does not imagine a dynamical attachment rule. 
The topology generator INET3.0~\cite{WinJam02} also falls
into this class. This generator has been specifically
designed to represent the Internet at the AS level by obtaining a
closely similar topology. As shown in Fig.~\ref{fig1}, such network
presents a small number of $k$-cores, with a shell size behavior that
is exponentially decreasing for increasing shell index.

Another Internet topology generator often discussed in the 
literature is the BRITE generator~\cite{BRITE}, which proposes a growth
mechanism combining the addition of vertices with $m$ new links according
to the preferential attachment with the addition of new links between
already existing vertices, also through a preferential attachment
mechanism. In this case, a non-trivial structure of shells is
obtained, with a largest shell index $k_{max}$ much larger than the
average degree, and a shell size decreasing as a power-law function of the
index. This implies a similar power-law relation between the size of
each $k$-core and its index, as observed in real Internet
maps. At large $k$, large fluctuations are
observed, with a relatively large central core (see
Fig.~\ref{fig1}). The difference between BRITE and BA 
networks highlights the structural relevance of the addition of new
links between already existing vertices in a growing heterogeneous
network model.

In general, as shown for an example in Fig.~\ref{fig1} (and with the exception
of the BA model), random networks with heavy-tailed degree distributions
present systematically a large number of shells (we have also checked that
$k_{max}$ increases if $\gamma$ decreases), and much larger than the average
degree $\langle d\rangle$. The shell size is decreasing as a power-law of the
index \cite{dorog06,dorog06b}, with a quite large central core of index
$k_{max}$, as for BRITE.  On the contrary, Weibull distributed networks have
relatively few shells with a much smaller $k_{max}$. It is interesting that
networks with relatively similar degree distributions can present in fact
strongly different $k$-core decompositions. This points to the $k$-core
decomposition as a supplementary valuable tool for network investigation.

\subsubsection{Core statistics and structure}

In this paragraph, we compare the characteristics
of the different cores, i.e. of more and
more central parts of the network. In the following we will focus only
on the models that have a core structure resembling that of the
Internet as the ER and the BA models are readily ruled out as possible
candidates to represent the Internet. 

\begin{figure}[t] 
\begin{center} 
\includegraphics[width=9cm,angle=0]{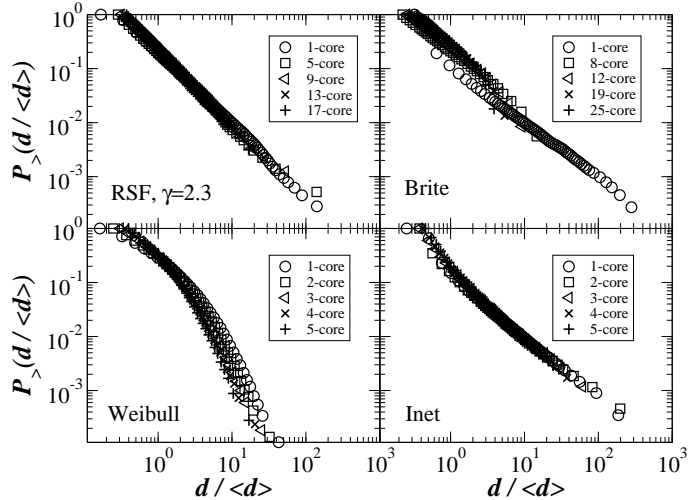} 
\end{center} 
\caption{Cumulative degree distribution of some $k$-cores for some
model networks. For each $k$-core, the degree is normalized by the
average degree of the core. For these various models, the collapse
of the various distributions show a striking
property of statistical self-similarity of the successive $k$-cores.
}
\label{Pk_models} 
\end{figure}

Figure~\ref{Pk_models} shows the cumulative degree distribution for some
$k$-cores, for some of the studied models; namely, the probability
$P_>(d)$ that any vertex in the networks has a degree larger than $d$.
Strikingly, the shape of the distribution (power-laws or Weibull) is
not affected by the decomposition. This feature, already noted
in \cite{dorog06} for uncorrelated scale-free networks, points to a striking
property of statistical self-similarity of the generated $k$-cores,
which resemble one with each other under the opportune rescaling of
the average degree.

\begin{figure}[thb] 
\begin{center} 
\includegraphics[width=90mm,angle=0]{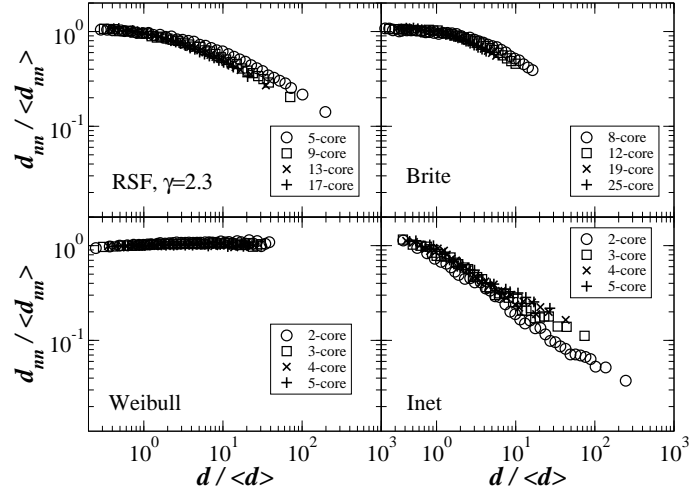} 
\end{center} 
\caption{Nearest neighbors degree distribution of some $k$-cores,
rescaled by the corresponding average values, for some model networks.
The degree of each node is normalized by the average degree of each
$k$-core. The data collapse confirms the statistical self-similarity of the 
cores.}
\label{Knn_models} 
\end{figure} 
 
\begin{figure}[thb] 
\begin{center} 
\includegraphics[width=90mm,angle=0]{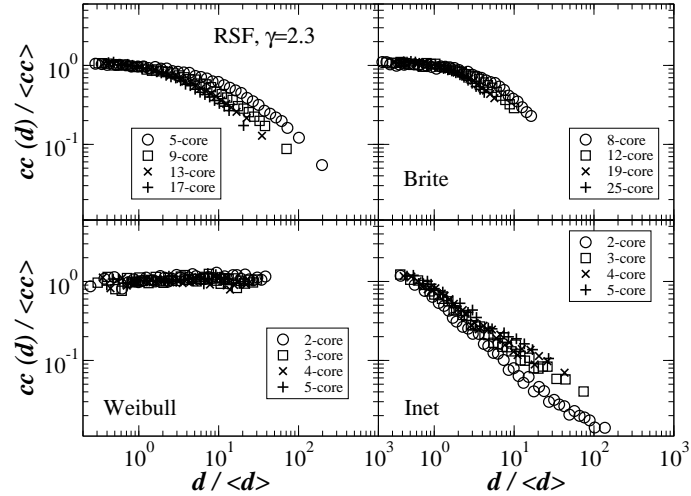} 
\end{center} 
\caption{Clustering coefficient spectrum of some $k$-cores 
for some model networks. The degree of each node
is normalized by the average degree of each $k$-core, and the clustering
coefficient is rescaled by the average clustering of each $k$-core. Once
again, a collapse is observed, confirming the self-similarity of the 
$k$-cores.
}
\label{Cnn_models} 
\end{figure}

As in the case of Internet maps, we characterize further this self-similarity
by computing the 2 and 3 point correlations as defined by the average degree
of nearest neighbors $d_{nn}(d)$ of vertices of degree $d$, and the clustering
spectrum $cc(d)$ of vertices of degree $d$. These quantities are reported in
Fig.s~\ref{Knn_models} and~\ref{Cnn_models} for the various
$k$-cores. Strikingly, the behavior of the two quantities is preserved
in all cases
as the network is recursively pruned of its low-degree vertices. In
other words, the overall network topology is invariant for $k$-cores
of increasing centrality.

\subsubsection{Summary}

In summary, the $k$-core decomposition allows to uncover
very different behaviors for different models which may otherwise
share e.g. very similar degree distributions. The $k$-core
decomposition is therefore a useful tool in the context of the 
model validation process. For example, 
a growing network obtained with the linear preferential attachment
rule may have a scale-free distribution of degrees $P(k)\sim k^{-\gamma}$
but will have a trivial shell structure because of its construction
mechanism. On the other hand, randomly constructed scale-free
networks, which may have weak correlation properties
and small clustering, can present a rich hierarchical 
decomposition with a large central core of high shell index. This
appears in agreement with the results of Ref.~\cite{dima} where
structural correlations and constraints appear to be sufficient to
determine most of the observed statistical properties observed in large
scale graphs. 

\section{
{\large $k$}-cores, dynamics and sampling biases}
\label{sec:dyn}
\subsection{Temporal variations of the $k$-core structure}

The availability of data obtained by the various projects makes it
possible to study the temporal evolution of the Internet maps. We have
considered the maps obtained by the CAIDA project at various times
between 2001 and 2005.  Table~\ref{table3} shows the main
characteristics of the analyzed maps, each of which was obtained from
the archives of one complete month.
\begin{table}[t]
\small
\begin{tabular}{|c|r|r|c|c|c|}
\hline
date & \# n & e & $\langle d\rangle$& $d_{max}$& $k_{\max}$ \\
\hline
\hline
 2001/05 & 7400  & 24791 & 6.700 & 1820 & 28 \\
 2002/03 & 8489  & 28871 & 6.802 & 2007 & 32 \\
 2003/05 & 8755  & 27300 & 6.236 & 1560 & 26 \\
 2004/04 & 9238  & 28016 & 6.065 & 1406 & 26 \\
 2005/04 & 8542  & 25492 & 5.969 & 1171 & 26 \\
\hline
\end{tabular}
\caption{Characteristics of the CAIDA
AS maps considered for the time analysis:
number of vertices $n$ and of edges $e$, average degree
$\langle d \rangle$, maximum degree $d_{max}$ and maximum shell
index $k_{\max}$.}
\label{table3}
\end{table}

While statistical signatures such as degree distribution,
disassortative behavior and clustering spectrum are typically very
stable over time, the $k$-core structure analysis 
reveals some finer variations.  For example, the number of vertices and
edges and the maximal shell index fluctuate in the CAIDA maps. This
can be tracked down to the fact that the number of sources used by
CAIDA changes (14 for the 2001/05 map, 21 for 2002/03, 24 for 2003/05
and 2004/04, and 22 for 2005/04), and that the locations of some of
these sources also change.

\begin{figure}[t] 
\begin{center} 
\includegraphics[width=90mm,angle=0]{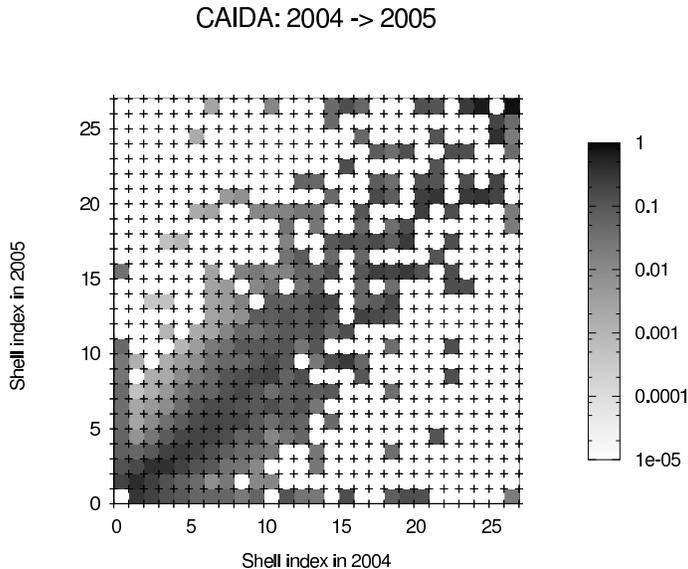}
\end{center} 
\caption{The grayscale code gives the probability of a change in shell
index, from the CAIDA map of 2004/04 ($x$ axis) to the one of 2005/04
($y$ axis).  The points in line $0$ correspond to ASes that are
present in 2004 but not in 2005, and the column $0$ corresponds to the
reverse situation. Most nodes do not change shell index, as the dark area
around the diagonal shows, but some important changes occur, with 
central nodes becoming peripherical, or vice-versa.}
\label{pchange} 
\end{figure}

Interesting informations also arise from the study of the change in
the composition of the various $k$-shells: we show as an example in
Fig.  \ref{pchange} the probability for a given AS to change from a
shell of index $x$ in a map obtained at a given time to a shell of
index $y$ in the successive map.  While most vertices do not change
their shell index, as shown by the dark area around the diagonal, some
suffer an important change of status, from a highly central shell to a
peripherical one or vice-versa.  This highlights the presence of strong
structural fluctuations in the evolution of CAIDA AS maps.

\begin{figure}[t] 
\begin{center} 
\includegraphics[width=90mm,angle=0]{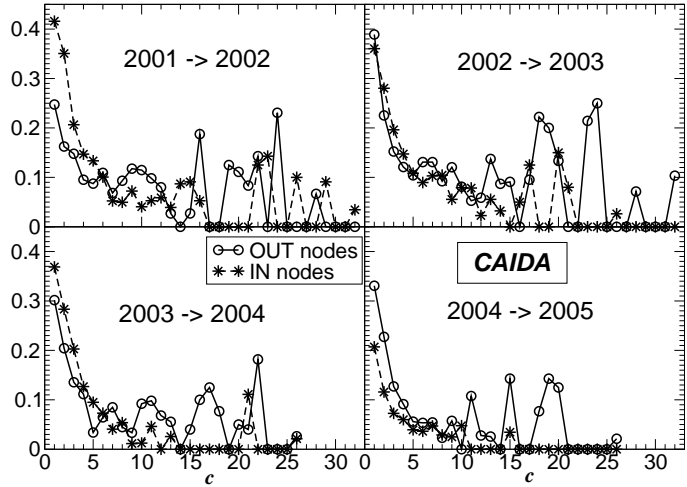}
\end{center} 
\caption{Probabilities that the vertices entering
(IN nodes) or disappearing from (OUT nodes) the CAIDA maps
have shell index $c$. Note how even nodes with large shell index
disappear from one year to the next.} 
\label{p_out_in} 
\end{figure} 

A further fingerprint of such structural changes is provided by the
analysis of the shell index of vertices that appear in or disappear
from the maps between one snapshot and the other, as shown in
Fig.\ref{p_out_in}: vertices in all shells, even central ones,
disappear from the CAIDA maps even in the most recent maps, between
2004 and 2005. The fluctuations observed in the shell index of ASes
may be related to three factors. A first one is the natural evolution of
the Internet structure. A second factor is the re-numbering of the ASes
for administrative reasons (see {\tt http://www.iana.org}).
A third factor is the uncertainty and bias in
the data collection. In this respect, CAIDA maps seem to exhibit a
high level of instability, indicative of a mapping process less stable
in time.  In this context, the $k$-core analysis appears as an
interesting tool to highlight the temporal changes of the Internet
structure as well as the measurement reliability in each particular
experimental set-up, at an intermediate level between global
quantities and local ones such as the degree.  It will certainly be of
interest in the future to study similar data for evolving DIMES maps,
which are obtained with a much larger set of sources.

\subsection{Sampling biases}

In this paragraph, we perform a sensitivity analysis of the $k$-core
decomposition with respect to potential sampling biases. In particular
we want to assess the effect of incompleteness and sampling biases on
the resulting structure of sampled graphs. For this reason we will
produce incomplete synthetic sampling processes of network models 
and compare the $k$-core structure of the sampled graph with that of
the original one. 

Internet maps are currently obtained through sampling methods of the
real Internet, which are based on a merging of paths between sources
and destinations, obtained either through Border Gateway Protocol
routing tables or through active \texttt{traceroute} measurements.
Such sampling processes present possible sources of errors and biases
whose effect has been up to now studied essentially for the degree
distributions~\cite{crovella,delos04,traceorsay,traceorsay2,claus05,latapy}.
The analysis of idealized sampling processes on networks with various
topologies has in particular revealed that the broadness of the degree
distributions observed in Internet maps is a genuine feature, although
important biases can remain on the exact form of the distribution, due
to an undersampling of vertices with small degree. Moreover, although a
path-based sampling process can produce a heterogeneous graph out of
an homogeneous initial network (such as an ER graph), as rigorously
shown in \cite{claus05}, this is restricted to the case of a single
source probing. It is therefore interesting to note that a single
source \texttt{traceroute}-like probing of any network yields
essentially a tree, whose $k$-core decomposition is by definition
trivial (with $k_{max}=1$). Another obvious but important confirmation
regards the largest shell index: by definition, a sampling cannot
discover paths or edges that do not exist, so that the maximal shell
index of a network, $k_{max}$, cannot be increased by partial sampling
(nor can the maximal degree observed). In fact the actual $k_{max}$ is
thus {\em at least} equal to the one found by a sampling of the true
network.

Since more central cores are more connected, and more paths go through
them, path-based sampling should intuitively discover and sample
better more central cores, while the peripherical shells could suffer
from stronger biases. In order to check such ideas, we perform a
\texttt{traceroute}-like probing of the various model networks considered
in section \ref{sec:models}, and compare their $k$-core decomposition before
and after sampling. We use the same model for \texttt{traceroute} as
in \cite{traceorsay,traceorsay2,latapy}: a set of $N_S$ sources sends probes to
$N_T$ destinations randomly placed on the network, and the shortest
paths between the source-destination pairs are merged to compose the
sampled network. We use $N_S=50$ sources, and various probing efforts
measured by $\epsilon=N_S N_T/N$ (where $N$ is the size of the initial
network), from a small value $\epsilon=0.1$ (corresponding to a small
density of targets $N_T/N=2. 10^{-3}$) to a much larger
$\epsilon=5$ (relatively large density of targets
$N_T/N=10^{-1}$).
\begin{figure}[thb] 
\begin{center}
\includegraphics[width=9.1cm,angle=0]{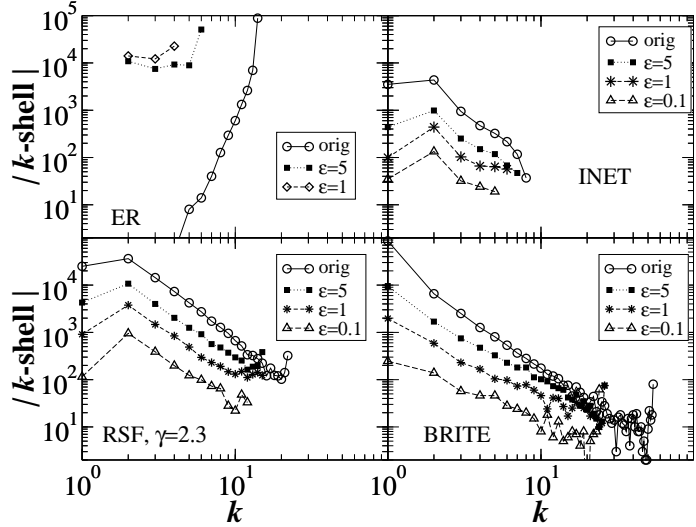}
\end{center}
\caption{Plot of the size of the $k$-shells vs. $k$ for various models,
before and after \texttt{traceroute}-like sampling, with different
probing efforts $\epsilon$. The qualitative shapes are
preserved by sampling.}
\label{biases_shell}
\end{figure} 

Figure~\ref{biases_shell} presents the curves of the $k$-shell size as
a function of the index for various network models and various
sampling efforts. For ER networks, the populated shells change from
being at index values only slightly under $k=\langle d \rangle$ to
much smaller values, with an almost uniform population of shells.  The
observed behavior is therefore completely different from the one
observed in AS maps. On the contrary, the power-law shape obtained
for RSF or BRITE networks, and comparable to the
one of the AS maps, is very robust, even if the slope is
affected. Indeed, shells of smaller indices are less well sampled. In
particular, the size of the first shell is most strongly decreased by
the sampling procedure; in some cases in fact, the first shell is
larger than the second in the original network, but becomes smaller in
the sampled network.  We note that in the available AS maps, the first
shell is indeed typically smaller than the second, and that the true
AS network thus very probably exhibits a much larger shell of index
$k=1$.  Similarly, one can expect that the exponent close to $2.7$ of
the power-law behavior of the shell size vs. its index (see
\cite{medusa2} and Fig.~\ref{shell_AS}) is a lower bound and that such
value might be reconsidered in the future thanks to more and more
extensive sampling efforts. On the other hand, the fact that the shell
of largest index is substantially larger than the ones with
immediately lower indices is well preserved, even if its index is
substantially decreased by the fact that many edges are ignored during
the sampling process.

\begin{figure}[thb] 
\begin{center} 
\includegraphics[width=90mm,angle=0]{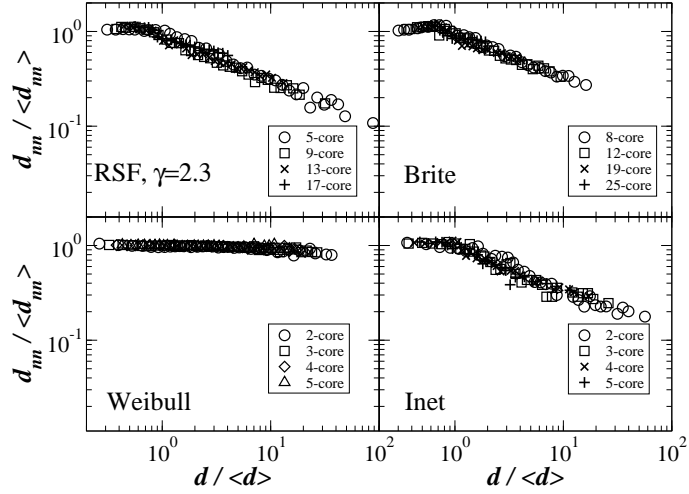} 
\end{center} 
\caption{Nearest neighbors degree distribution of some $k$-cores,
rescaled by the corresponding average values, for some network models
after sampling through a \texttt{traceroute}-like process
with $N_S=50$ sources and target density $N_T/N=0.1$. The data collapse
shows that the self-similarity
is preserved by sampling.}
\label{Knn_TR} 
\end{figure}

\begin{figure}[thb] 
\begin{center} 
\includegraphics[width=90mm,angle=0]{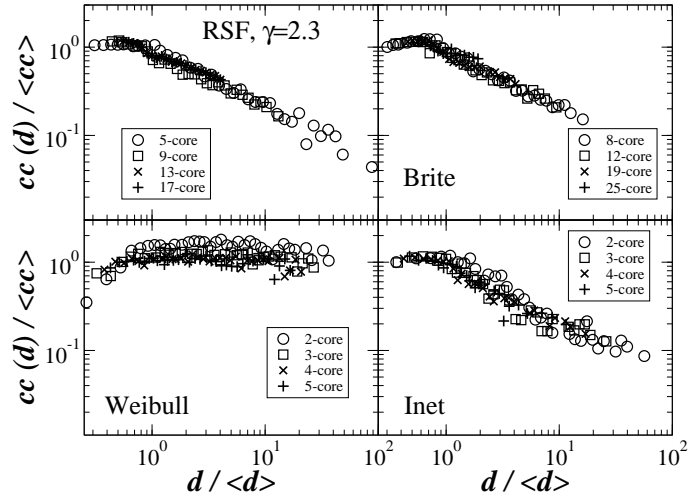}
\end{center} 
\caption{Clustering spectrum of some $k$-cores,
rescaled by the corresponding average values, for some network models
after sampling through a \texttt{traceroute}-like process
with $N_S=50$ sources and target density $N_T/N=0.1$.}
\label{Cnn_TR} 
\end{figure}

Figures \ref{Knn_TR} and \ref{Cnn_TR} moreover show that
the self-similar properties of the $k$-core decomposition are
preserved by the sampling process. Although the precise form of the
degree distribution of the whole network is slightly altered, the
basic correlation properties are conserved by the sampling. Moreover,
the self-similar structure of the $k$-core decomposition is also
preserved, as a comparison of Fig.s~\ref{Knn_TR} and
\ref{Cnn_TR} with Fig.s~\ref{Knn_models} and
\ref{Cnn_models} clearly shows.

\begin{figure}[thb]
\vskip -.5in
\begin{center}
\includegraphics[width=9.cm,angle=0]{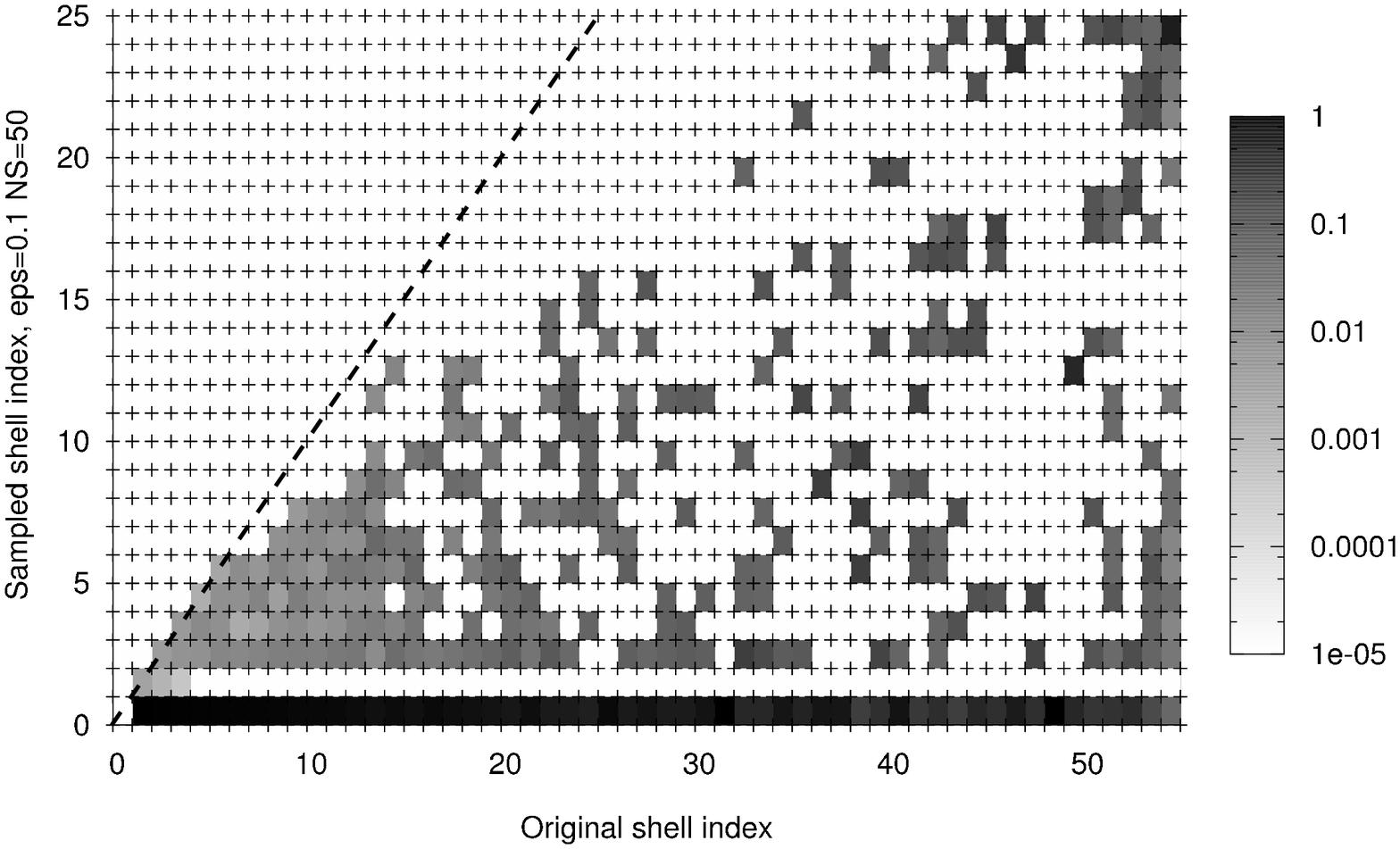}
\end{center}
\caption{The grayscale code gives the probability of a change in shell
index due to the \texttt{traceroute}-like sampling, from a certain
index before sampling ($x$ axis) to another one after sampling ($y$
axis).  The line at $y=0$ represents the probability of vertices of shell
index $x$ to be absent from the sampled graph. The initial network is
obtained by the BRITE generator. Here $N_S=50$ sources and a fraction
$N_T/N=2. 10^{-3}$ of targets are used. The low sampling effort
implies that many nodes are not discovered, and that the measured
shell index can differ strongly from the original one.}
\label{diff.1}
\end{figure}

\begin{figure}[thb]
\vskip -.5in
\begin{center}
\includegraphics[width=9.cm,angle=0]{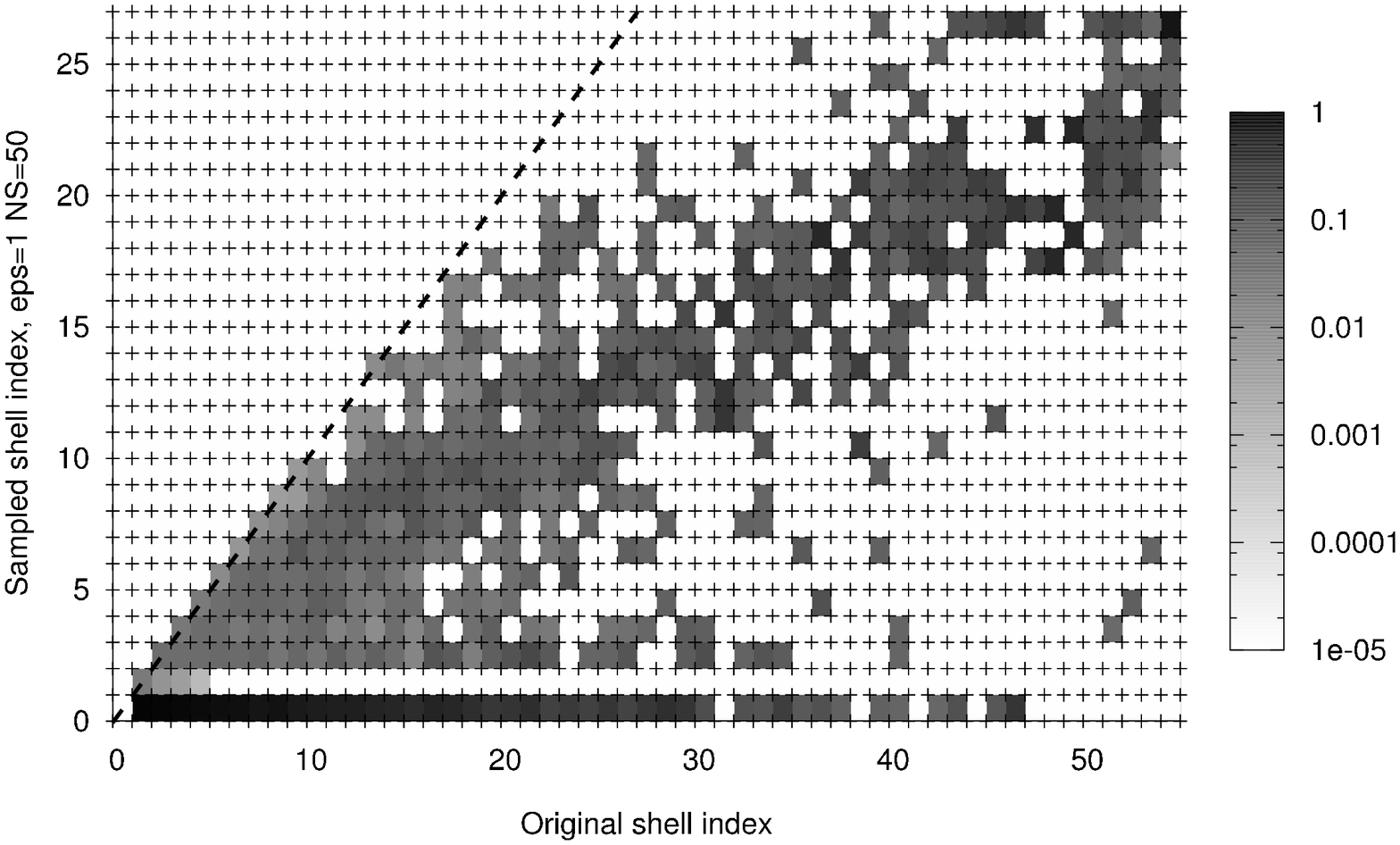}
\vskip -.2in
\includegraphics[width=9.cm,angle=0]{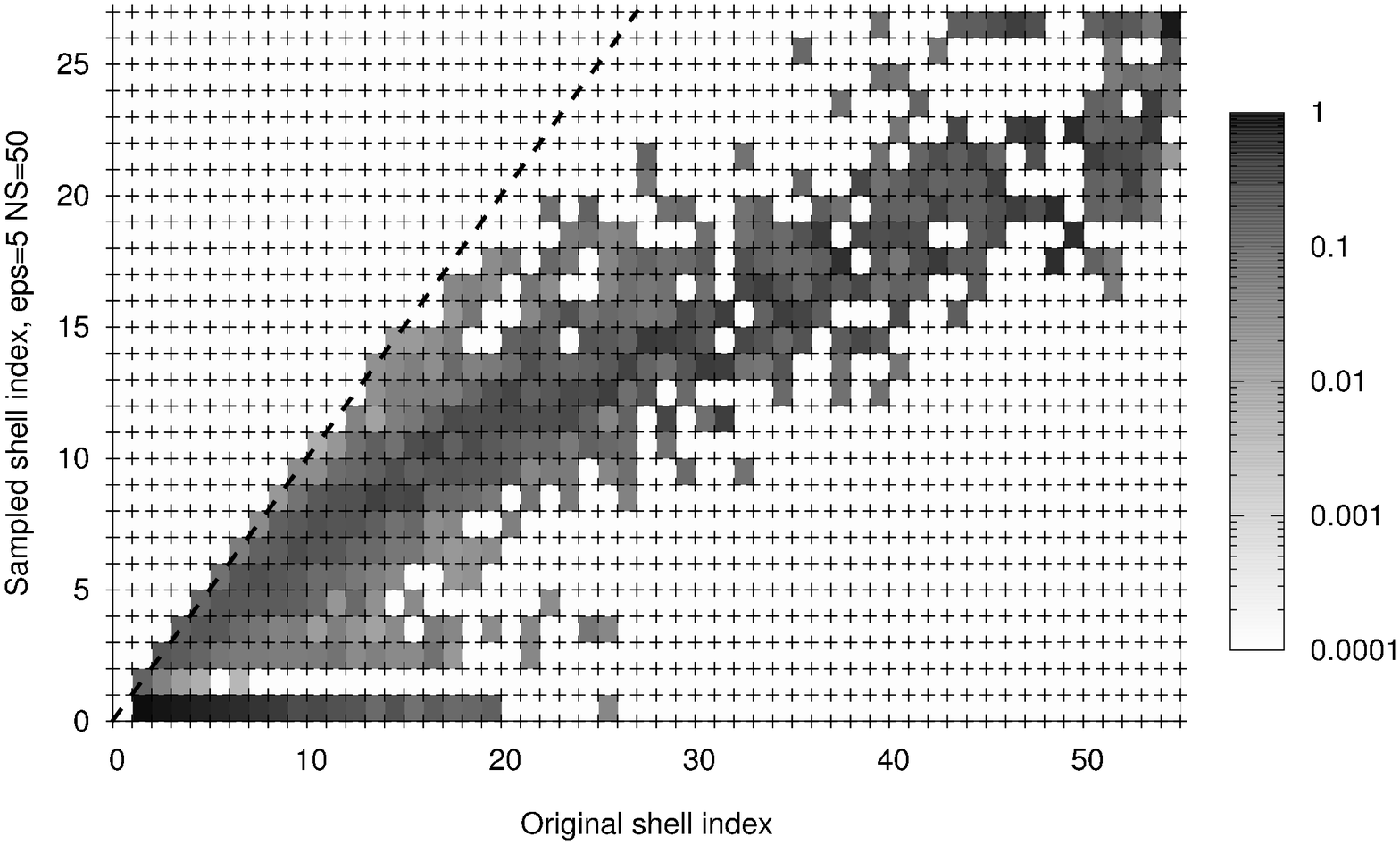}
\end{center}
\caption{Same as Fig.~\ref{diff.1} for $N_S=50$ sources and
$N_T/N=2. 10^{-2}$ (top) and $N_T/N=10^{-1}$ (bottom). As the sampling effort
is increased, the measured and the original shell index become more 
correlated.}
\label{diff1}
\end{figure}

While the main statistical properties of the $k$-core decomposition
are therefore largely conserved by the sampling process, allowing to
distinguish between networks with different topological structures,
important quantitative biases can appear and compromise the accuracy
of the measurements, as we now investigate.  In order to understand
such effects in more details, we indeed show in Fig.s~\ref{diff.1} and
\ref{diff1} the probability for a vertex of given shell index in the
original network to have another shell index in the sampled network,
in the case of an original network obtained by the BRITE generator. At
low sampling effort, many vertices are simply left undiscovered, and
the shell index properties can be strongly affected in a seemingly erratic
way, as shown by the important scattering of data in
Fig.~\ref{diff.1}. As soon however as the sampling effort is increased
to a more reasonable level, a strong correlation appears between the
true shell index and its value in the sampled graph, even if a
systematic downwards trend is observed (Fig. \ref{diff1}).

In summary, our results indicate that the sampling biases do in fact
affect only slightly the measure of the statistical properties of
heterogeneous graphs and of their $k$-core decomposition, even at
relatively low level of sampling. In fact, the routing properties as
``measured'' by the shell indices will be in fact rather
underevaluated due to the incomplete sampling of edges, which can be
taken as a rather good news showing that the AS network probably
offers better performance (QoS, robustness) than what can be
measured by the present maps.

\section{Conclusions}\label{concl} 
 
We have presented the application of the $k$-core decomposition to the
analysis of large scale networks models and of large scale Internet
maps.  The $k$-core decomposition allows the progressive pruning of
the networks and the identification of subgraphs of increasing
centrality. These subgraphs have the property of being more and more
densely connected, and therefore of presenting more and more robust
routing capabilities.  The study of the obtained subgraphs uncovers
the main hierarchical layers of the network and allows for their
statistical characterization. Strikingly, we observe 
for the Internet at the Autonomous System a
statistical self-similarity of the topological properties for cores of
increasing centrality.

The $k$-core decomposition proves useful to uncover not only the
hierarchical decomposition of real maps, but also for model
validations. For example, many models, although having, {\em e.g.}
degree distribution and clustering properties similar to those of real
maps, do not present shell index values as large as the real data, nor
a similar structure in which each $k$-core is composed by a constant
fraction of the $k-1$-core. The $k$-core decomposition should therefore be
considered as a {\em supplementary} valuable tool for network
characterization and model validation.

It is also worth mentioning that the router level $k$-core structure
of the Internet appears to have different properties than those
appearing at the AS level~\cite{DIMES,LANET-VI}. This calls for
repeating the present analysis for different router level maps
available at the moment in order to better emphasize the structural
difference exhibited by the two different mapping granularities.

Moreover, the $k$-core analysis allows to compare maps obtained by 
different mapping processes, follow their temporal evolution and
assess the stability of these maps. It also appears as an interesting way of
discriminating between various topologies, even after sampling
biases have been introduced: for example, a sampled ER network
may display a power-law like degree distribution in case of a very 
limited sampling effort, but its $k$-core decomposition will in any case
remain very different from the one of sampled heterogeneous networks.

Finally, the $k$-core decomposition may be used also to define a
computational feasible centrality measure and a hierarchy between the
nodes of a network. It combines the degree ranking with more global
structural properties, connectedness and routing capabilities, providing
a centrality measure that is highly correlated with the various
standard definitions such as degree and betweenness centrality.

In conclusion, the $k$-core decomposition appears at a general level
as a very interesting and useful additional tool for analysis of complex
networks, with particular relevance in the context of technological and
communication networks.

\section*{Acknowledgments}
This work has been partially supported by the EU within the 6th Framework
Program under contract 001907 (DELIS).
We acknowledge also interesting discussions with
Pierre Fraigniaud. A.V. is partially funded by the NSF IIS-0513650
award.

\medskip

\end{document}